\documentstyle[11pt]{article}
\renewcommand{\baselinestretch}{1.24}
\newcommand{\NP}[1]{Nucl.\ Phys.\ {\bf #1}}
\newcommand{\PL}[1]{Phys.\ Lett.\ {\bf #1}}

\newcommand{\PR}[1]{Phys.\ Rev.\ {\bf #1}}

\newcommand{\MPL}[1]{Mod.\ Phys.\ Lett.\ {\bf #1}}
\newcommand{\IJMP}[1]{Int.\ J.\ Mod.\ Phys.\ {\bf #1}}

\newcommand{\be}{\begin{equation}}
\newcommand{\en}{\end{equation}}
\newcommand{\bea}{\begin{eqnarray}}
\newcommand{\ena}{\end{eqnarray}}
\newcommand{\W}{{\sf W}}
\newcommand{\WA}{{\sf WA}}

\newcommand{\WD}{{\sf WD}}
\newcommand{\wh}[1]{\widehat{#1}}
\def\BRT{{\rm I{\hbox to 5.5pt{\hss\rm R}}}}
\def\BRS{{\mbox{\hbox{\small I{\hbox to 4.3pt{\hss\small R}}}}}}
\def\BRSS{{\mbox{\hbox{\footnotesize I{\hbox to 3.35pt{\hss\footnotesize
R}}}}}}

\def\BZT{{\rm Z{\hbox to 3pt{\hss\rm Z}}}}
\def\BZS{{\mbox{\hbox{\small Z{\hbox to 2.3pt{\hss\small Z}}}}}}
\def\BZSS{{\mbox{\hbox{\footnotesize Z{\hbox to 1.8pt{\hss\footnotesize Z}}}}}}

\newcommand{\trunc}{\triangleright}

\newcommand{\vs}[1]{\rule[ - #1 mm]{0 mm}{#1 mm}}
\newcommand{\sect}[1]{\setcounter{equation}{0}\section{#1}}

\begin{document}
\setcounter{page}{0}
\pagestyle{empty}
\renewcommand{\thefootnote}{\fnsymbol{footnote}}
%
%      Include here the text for the titlepage:
%
\rightline{DFTT-40/94}
\rightline{hep-th/9410013}
\rightline{September 1994}

\vs{15}

\begin{center}
{
\LARGE { \W-algebras of negative rank }
 }\\[1cm]
{\large K.\ Hornfeck}\\[0.5cm]
{\em INFN, Sezione di
Torino, Via Pietro Giuria 1, I-10125 Torino,
Italy}
\\[1cm]
\end{center}
\vs{15}
\centerline{\bf Abstract}
Recently it has been discovered that the \W-algebras (orbifold of) \WD$_n$ can
be defined even for negative integers $n$ by an analytic continuation of their
coupling constants. In this letter we shall argue that
also the
algebras \WA$_{-n-1}$ can be defined and are finitely generated. In addition,
we
show that  a
surprising connection exists between already known \W-algebras, for example
between the {\sf
CP}$(k)$-models and the {\sf U}$(1)$-cosets of the
generalized Polyakov-Bershadsky-algebras.
\renewcommand{\thefootnote}{\arabic{footnote}}
\setcounter{footnote}{0}
\newpage
\pagestyle{plain}
\sect{Introduction}
Nowadays, the Casimir algebras like \WA$_{n-1}$ or \WD$_n$ are well-understood.
They are connected to the classical Lie-algebras {\sf sl}$(n)$ and {\sf
so}$(2n)$~\cite{FL90,BS93}. In~\cite{Hor94} it was argued that a special
algebra of type
\W$(2,4,6)$~\cite{KW91} can be
regarded as the algebra \WD$_{-1}$ in the sense that its coupling constants
are given by those of the orbifolds of the algebras \WD$_n$ by inserting
$n=-1$.
Moreover, on the basis of the minimal models for \WD$_n$~\cite{LF89}
 and a small set of known
exact examples~\cite{EHH93} a conjecture on the minimal models of this algebra
could be stated that has been proven in~\cite{BEHal94a} to be correct
by identifying this
algebra with an {\sf sl}$(2)$-coset. Whereas in ref.~\cite{Hor94} the existence
of this
\WD$_{-1}$-algebra was just an amazing coincidence,
in~\cite{BEHal94a} the existence of algebras \WD$_n$ for all negative integers
$n$ has been conjectured and their field content, their minimal
models  as well as a coset-realization was
given. In this letter we want to extend
the notion of \W-algebras of negative rank to algebras of type
\W$(2,3,\ldots,N)$ that are so-called ``unifying algebras'' for
\WA~\cite{BEHal94a,BEHal94} (see section~3). These include the algebras
\WA$_{n-1}$ themselves,
but also the {\sf CP}$(n)$-models and the cosets of the generalized
Polyakov-Bershadsky- algebras (P-B-algebras) \W$^{\mbox{\sf \small
sl}(n)}_{\mbox{\sf \small
sl}(n-1)}$/{\sf
U}$(1)$ and others. One should note at
this point that the expression ``negative rank''
refers to the underlying Lie-algebra (in this case {\sf sl}$(n)$)
and not to the field-content
of the resulting \W-algebra that we shall denote for example
 as \WA$_{-n-1}$ or {\sf
CP}$(-n)$. These \W-algebras will again be of type
\W$(2,3,\ldots,N)$, where however the functional dependence of $N$ (the
highest
dimension of the simple fields in the \W-algebra) on $n$ differs, whether $n$
is
positive or negative. In fact, for most of these \W-algebras of negative rank
we shall recover other well-known \W-algebras and in this way reveal a close
relationship between at first glance completely different \W-algebras. The only
exception to this are the \WA$_{-n-1}$-algebras that are so far unknown
algebras.
\sect{Universal \W-algebras}
Universal \W-algebras are algebras, in which the coupling constants depend on
(at least) one additional {\em free}
parameter apart from the central
charge~$c$~\footnote{Note that our ``universal algebras'' are different from
those of ref.~\cite{BK90}.}. By fixing this additional parameter to different
values we can
describe very different algebras. A simple example is the algebra \W(2,2), in
which the selfcoupling ${\cal C}_{22}^2$ remains completely undetermined.

More interesting examples are \W-algebras of cosets like
\be
\frac{\mbox{\sf \small sl}(n)_\kappa \oplus \mbox{\sf \small sl}(n)_\mu}
{ \mbox{\sf \small sl}(n)_{\kappa+\mu}} \ \ .
\label{cos}
\en
These algebras soon become
very complicated, their field content for the first $n$
is~\cite{Blu91,Vos92}
\bea
&n=2:&\mbox{\W}(2,4,6^2,8^2,9,10^2,12) \ \ ,\nonumber \\[1mm]
&n=3:&\mbox{\W}(2,3,4,5,6^4,7^2,8^7,9^9,10^{12},11^{16},12^{26},13^{26},
14^{33},15^{33},
16^{12}) \ \ ,\label{22} \\[1mm]
&n=4:&\mbox{\W}(
2,3,4^2,5,6^5,7^4,8^{12},9^{15},10^{29},11^{42},12^{80},13^{111},14^{191},
15^{286},
16^{435}, \nonumber \\
&& \hspace{1cm} 17^{613},18^{857},19^{1054},20^{1140},21^{743})\ \ . \nonumber
\ena
One of the levels, say $\mu$, can be
replaced by the central charge
\be
c \,=\, \frac{\kappa\,\mu\,(n^2-1)\,(\kappa+\mu+2n)}{(\kappa+n)\,(\mu+n)\,
(\kappa+
\mu+n)} \ \ ,
\label{ccc}
\en
the other one, $\kappa$, is open and describes different algebras, for example
$\kappa=1$ gives the algebra \WA$_{n-1}$~\cite{BBSS88a}. This shows that
for some special values of $\kappa$ the algebra truncates to a simpler one.
The case $n=2$
has been treated in detail in~\cite{Blu91,BEHal94a} and we give only the
result: Truncations of the generic algebra take place
for positive
integers $\kappa$ less than 6~\cite{Blu91}, as well as for a couple of negative
rational numbers.
The complete set of truncations up to dimension~9
has been given in~\cite{BEHal94a}:
\begin{center}
\begin{tabular}{||c|c||}         \hline
$\kappa$ & {\em algebra} \\[2mm] \hline
{\em generic} & \W$(2,4,6^2,8^2,9,10^2,12)$ \\
1 & \W$(2)$    \\
2 & \W$(2,4,6)$ \\
3 & \W$(2,4,6^2,8,9)$ \\
4 & \W$(2,4,6^2,8^2,9,10)$ \\
5 & \W$(2,4,6^2,8^2,9,10^2)$ \\
$-\frac{1}{2}$  &  \W$(2,4,6)$ \\
$-\frac{4}{3}$  &  \W$(2,6,8,10,12)$ \\
$-\frac{8}{5}$  &  \W$(2,4,6,8,9,10,12)$ \\
$-\frac{12}{7}$ &  \W$(2,4,6^2,8,9,10)$ \\ \hline
\end{tabular}
\end{center}
\centerline{Table I: Truncations of the coset-algebra~(\ref{cos}) for $n=2$}
As we see, the ``universal'' coset-algebra in this case
can even describe
\W-algebras of different field content. The two algebras in this set of type
\W$(2,4,6)$ ($\kappa=2$ and $\kappa=-1/2$) are not identical; the first one
is the bosonic projection of the super-Virasoro algebra the other with
different coupling constants~\cite{KW91} and~minimal models~\cite{EHH93} is
the special algebra that
we called in the introduction the algebra \WD$_{-1}$.
Nevertheless, the coupling constants for the coset can be (in principle)
determined as function of $c$ (replacing the level $\mu$)
and the parameter $\kappa$, for
example~\cite{BEHal94a}
\bea
\left({\cal C}_{44}^4(\kappa,c)\right)^2 & = & \label{scsl2cos}
\\[1mm]
&&\hspace{-3cm}
{{3\,{{\left( -496\,c + 40\,{c^2} - 672\,\kappa + 536\,c\,\kappa +
160\,{c^2}\,\kappa -
          1416\,{\kappa^2} + 844\,c\,{\kappa^2} + 140\,{c^2}\,{\kappa^2} -
354\,{\kappa^3} +
          211\,c\,{\kappa^3} + 35\,{c^2}\,{\kappa^3} \right) }^2}}\over
   {5\,\left( 22 + 5\,c \right) \,\left( -1 + \kappa \right) \,
     \left( 4 + 3\,\kappa \right) \,\left( 6\,c - 5\,\kappa + 5\,c\,\kappa -
{\kappa^2} +
       c\,{\kappa^2} \right) \,\left( 4\,c + 48\,\kappa + 8\,c\,\kappa +
18\,{\kappa^2} +
       3\,c\,{\kappa^2} \right) }}\ . \nonumber
\ena

Another interesting universal algebra is the one that contains {\em all}
 the
algebras \WA$_{n-1}$ and I shall denote it therefore \WA$^{n-1}$. Note the
difference between the two algebras: \WA$_{n-1}$ is only defined for
positive
integer $n$ and is an algebra of type \W$(2,3,\ldots,n)$, whereas the
universal algebra \WA$^{n-1}$ is a generic algebra of \W$_{\infty}=
\W(2,3,\ldots,\infty)$-type with a free parameter $n$ that is a real
number and that plays the r\^ole of $\kappa$ in the
coset~(\ref{cos})~\footnote{By \W$_{\infty}$ (\W$_{1+\infty}$)
we denote the algebra
that has a linear basis. Other algebras with an infinite set of simple fields
we shall call ``of \W$_{\infty}$-type''.}.
Of course, whenever in the universal algebra \WA$^{n-1}$ the parameter
$n$ is a positive integer, the coupling constants of the universal algebra
should coincide with the coupling constants of \WA$_{n-1}$ and the algebra
should truncate
\be
\mbox{\WA}^{n-1} \, \trunc \, \mbox{\WA}_{n-1}\hspace{2cm} \mbox{for $n$
positive integer}.
\label{t1}
\en

The first coupling constants (in
standard normalization and basis for
the fields) for this universal algebras are
known~\cite{Hor92,BEHal94a} and we cite as an illustration just the coupling
constants $\left({\cal C}_{33}^4\right)^2$ and $\left({\cal C}_{44}^4
\right)^2$
(where the relative sign between ${\cal C}_{33}^4$ and ${\cal C}_{44}^4$ is
fixed):
\bea
\left({\cal C}_{33}^4[{\mbox{\WA$^{n-1}$}}]\right)^2 & = &
 \frac{64 \,(n-3)\,(c+2)\,(c\,(n+3) + 2 \,
(4n+3)\,(n-1))}{(n-2)\,(5c+22)\,(c\,(n+2) + (3n+2)\,(n-1))} \ \ ,
\label{4} \\[2mm]
\left({\cal C}_{44}^4[{\mbox{\WA$^{n-1}$}}]\right)^2 & = &\label{4a}
 \\[1mm]
&& \hspace{-3cm}
{{36\,{{\left( 82 + 45\,c - 19\,{c^2} - 94\,{n^2} - 75\,c\,{n^2} +
          {c^2}\,{n^2} + 12\,{n^3} + 18\,c\,{n^3} \right) }^2}}\over
   {\left( 2 + c \right) \,\left( 22 + 5\,c \right) \,\left( -3 + n \right) \,
     \left( -2 + n \right) \,\left( -2 + 2\,c - n + c\,n + 3\,{n^2} \right) \,
     \left( -6 + 3\,c - 2\,n + c\,n + 8\,{n^2} \right) }} \ \ .\nonumber
\ena
For positive integer values for $n$, these coupling constants
indeed
reproduce those of \WA$_{n-1}$. In addition, also the two infinite algebras
\W$_{\infty}$ and \W$_{1 + \infty}$/{\sf U}$(1)$ are contained
in the universal algebra
as can be checked immediately by comparing the coupling constants in
the basis of primary fields
\be
\mbox{\WA}^{\infty} \, = \, \mbox{\WA}^{0} \, = \, \mbox{\WA}^{-2} \, = \,
\mbox{\W}_{\infty} \hspace{1cm};\hspace{1cm}
\mbox{\WA}^{-1} \,=\, \frac{\mbox{\W}_{1+\infty} }{\mbox{\sf U}(1)} \ \ .
\en

As we mentioned, the algebra \WA$^{n-1}$ is for general values of $n$ an
algebra of \W$_{\infty}$-type and it truncates for positive integers $n$ to
\W$(2,3,\ldots,n)$. We can see from the coupling constants~(\ref{4})
and~(\ref{4a}) how this truncation works in the case $n=3$. The
coupling constant ${\cal C}_{44}^4$ shows a pole at $n=3$. To render
the coupling
constants well-defined, we have to renormalize the field $W_4$ (and in
fact all fields $W_k$ with $k \geq 4$) to
\be
\widetilde{W}_k \, = \, W_k \,\sqrt{n-3}\hspace{1.5cm}
\mbox{for all $k
\geq 4$.}
\label{W4renorm}
\en
In the new fields all coupling constants are well-defined at $n=3$,
but all central terms except those in the OPEs of $T \star T$ and $W_3 \star
W_3$ vanish. Therefore all fields $\widetilde{W}_k$, $k\geq 4$, will be
nullfields and can be dropped from the algebra and \WA$^2$ truncates to \WA$_2
= $ \W$(2,3)$.

In the following we shall no longer distinguish between the universal algebra
\WA$^{n-1}$ and \WA$_{n-1}$ and use only the notation \WA$_{n-1}$, having in
mind that it is defined for real $n$ and is for generic $n$ an algebra of
\W$_{\infty}$-type.
Later we shall argue that the algebra \WA$_{n-1}$ dos not only truncate
for positive, but also for negative integers $-n$, a \WA-algebra of negative
rank. The resulting \W-algebra
will be of type \W$(2,3,\ldots,(n+1)^2-1)$.

Up to this point \WA$_{-n-1}$ is just
defined by the
continuation of the coupling constants of \WA$_{n-1}$ to negative
$n$. In the last section we shall give a coset-construction of the algebras
\WA$_{-n-1}$ that is more inspired by the classical Lie-algebras. We would also
like to
note at this point that the highest dimension of the simple fields in the
\W-algebra goes no longer linearly with $n$, but quadratically. This was also
observed in~\cite{BEHal94a} in the case of the algebra \WD$_{-n}$.
Here we collect the field content of \WA$_{n-1}$ for all integers $n$:
\begin{center}
\begin{tabular}{|c|c|}\hline
$n$ & \WA$_{n-1}$ \\[2mm] \hline
-1,\,1,\,$\pm \infty$ & \W$_\infty$ \\[1mm]
0 & \W$_{1 + \infty}$/{\sf U}$(1)$ \\[1mm]
$n\geq 2$, {\em integer} & \W$(2,3,\ldots,n)$ \\[2mm]
$n\leq-2$, {\em integer} & \W$(2,3,\ldots,(|n|+1)^2-1)$ \\ \hline
\end{tabular}
\end{center}
\centerline{Table II: Field content of \WA$_{n-1}$ for integer $n$}
\sect{Unifying \W-algebras}
In~\cite{BEHal94a,BEHal94} it has been shown that the algebra \WA$_{n-1}$
truncates at special values of the central charge
to an algebra with less fields (if $n$ is sufficiently large), especially
at all of their minimal models. Since these truncated algebras are the same for
all $n$, they are called ``unifying algebras''. For example, the $k$th unitary
model of \WA$_{n-1}$ is via level-rank duality~\cite{BG88,Alt89}
described by the so-called {\sf CP}$(k)$-model~\cite{BS93,BK90a,Nar91,BK92}
\be
\mbox{\sf CP}(k) \, \cong \, \frac{\mbox{\sf sl}(k+1)_n}{\mbox{\sf
sl}(k)_n\otimes \mbox{\sf U}(1)} \ \ .
\label{cpk}
\en
In~\cite{BFH94} it has been argued that the {\sf CP}$(1)$-model is a
\W-algebra of type \W$(2,3,4,5)$ and in~\cite{BEHal94a} it has been conjectured
that the general {\sf CP}$(k)$-model is of type
\W$(2,3,\ldots,k^2 + 3 k +1)$.

In general, if we parameterize the central charge by
\be
c(p,q) \, = \, (n-1) \, \left(1 - n\,(n+1)\,\frac{(p-q)^2}{p\,q}\right) \ \ ,
\label{cdeg}
\en
the algebra \WA$_{n-1}$ truncates to the unifying algebra \W$_{\{r,s\}} = $
\W$_{\{s,r\}} = $
\W$(2,3,\ldots, r s -1)$
at the central
charge
\be
c \,= \,c_1^{r,s}(n) \,=\, c(n-1+r, n-1+s) \mbox{\hspace{2cm}
for $r\neq s$.}
\label{1}
\en

The unifying algebras \W$_{\{r,1\}}$ are the \WA$_{r-2}$ themselves, the
\W$_{\{r,2\}}$ are the cosets of the generalized
P-B-algebras,
\W$^{\mbox{\sf \small sl}(r-1)}_{\mbox{\sf \small sl}(r-2)}$.
Here we use for a \W-algebra of the DS-reduction of an
{\sf sl}$(2)$ embedding $\cal S $ into ${\cal G}$ (in our case ${\cal G} =
${\sf
sl}$(r-1)$) the notation \W$^{\cal G}_{\cal S}$.
A proposal~\cite{Rag94}, checked for a variety of cases in
ref.~\cite{BEHal94a}, states that the general unifying
\W-algebra
\W$_{\{r,s\}}$, $s < r$, is given by the coset
\be
\mbox{\W}_{\{r,s\}} \,=\,\frac{
\mbox{\W}^{\mbox{\sf \small sl}(r-1)}_{\mbox{\sf \small
sl}(r-s)}}{\mbox{Kac-Moody
subalgebra}}\ \ ,
\label{UDS}
\en
where the KM-subalgebra is a {\sf sl}$(s-1)\otimes ${\sf U}$(1)$ for
$s \geq 2$, whereas for $s=1$ no such subalgebra is present (and hence one has
not to take a coset). In the ``limit'' $s=r-1$ one recovers the {\sf
CP}$(r-2)$-model.

Due to the symmetries of the coupling constants there are in general
two
other
truncations at
\bea
c &=& c_2^{r,s}(n) \,=\, c(n+1-r, n+s-r) \mbox{\hspace{2cm}
for $s \neq 1$,}                          \label{2} \\
c &=& c_3^{r,s}(n) \,=\, c(n+1-s, n+r-s) \mbox{\hspace{2cm}
for $r \neq 1$.}
\label{3}
\ena
The restrictions in eqs.~(\ref{1}),~(\ref{2}) and~(\ref{3}) come from the fact
that the
derivation of the truncation fails whenever $p=q$.

One important consequence of the truncations is that one can recover the
coupling constants for {\em all} algebras \W$_{\{r,s\}}$, if they are known
for
a series, i.e.\ from the coupling constant ${\cal C}[\mbox{\W}_{\{r,s_o\}}]$
with a fixed $s_o$ one can compute the same coupling constant for all the
other $s$. Therefore one can derive the coupling constants
of the algebras \W$_{\{r,s\}}$ from those of \WA$_{n-1}$ by solving for example
$c
= c_1^{r,s}(n)$ for $n$ and inserting it into the coupling constants for
\WA$_{n-1}$~(\ref{4}),~(\ref{4a}).
The result is independent of the choice $c_1^{r,s}$,
$c_2^{r,s}$ or $c_3^{r,s}$. These functions therefore yield a {\em
parameterization\/} of the coupling constants of \W$_{\{r,s\}}$. Since
the
solution for the coupling constants show in general roots when expressed in
the central charge, it is
often
more convenient to leave the coupling constants in the parameterization $c(n)$.

As an exercise let us compute the coupling constant ${\cal C}_{33}^4$ for
the {\sf CP}$(1)$-model (\W$_{\{3,2\}}$).
We have to solve
\be
c \, = \, c_1^{3,2}(n) \, = \, \frac{2 \, (n-1)}{n+2}
\label{7}
\en
for $n$ and insert it into eq.~(\ref{4}).
We obtain immediately
\be
\left({\cal C}_{33}^4[\mbox{{\sf CP}$(1)$}]\right)^2 \,=\,
\frac{16}{3}\,\frac{(c+2)\,(c+10)^2\,(5c-4)}{(c+7)\,(2c-1)\,(5c+22)}
\label{8}
\en
that is the correct coupling constant in standard
normalization~\cite{Nar91,BK92,Hor93}. In the same way we can treat the coset
of the P-B-algebra (\W$_{\{4,2\}}$). We have the parameterization
\be
c \, = \, c_1^{4,2}(n) \, = \, - 3 \, \frac{(n-1)^2}{n+3}
\label{9}
\en
%leading to the coupling constant
%\be
%\left({\cal C}_{33}^4[\mbox{\W}^{\mbox{\sf \small sl}(3)}_{\mbox{\sf \small
%sl}(2)}]\right)^2
%\,=\,
%{{16\,{{\left( 3 - n \right) }^2}\,\left( 3 + n \right) \,
%     \left( 1 + 3\,n \right) \,\left( 9 + 5\,n \right) }\over
%   {\left( 2 - n \right) \,\left( 3 + 2\,n \right) \,
%     \left( 51 + 52\,n - 15\,{n^2} \right) }}
%\label{10}
%\en
and by defining $n = 2k+3 $ we recover the coupling constant~(2.1.33)
of ref.~\cite{BEHal94a}.
It is an easy exercise to check that the parameterizations $c_2^{r,s}(n)$ and
$c_3^{r,s}(n)$ give the same coupling constants.
\sect{\W-algebras of negative rank}
The existence of the unifying algebras \W$_{\{r,s\}}$ has been proven only for
$s\neq r$. Also the conjecture~(\ref{UDS}) is only valid in this case, as is
the parameterization $c_1$. On the other hand, however, the parameterizations
$c_2$ and $c_3$ are well-defined even in the case $s=r$ and the truncation of
\W$_{1+\infty}$/{\sf U}$(1)$ at $c=-r$ to an algebra of type
\W$(2,3,\ldots,r^2-1)$ (following from~\cite{AFMO94}) suggests the existence of
an unifying
algebra
\W$_{\{r,r\}}$.

According to the last section, we can compute the structure
constant
$\left({\cal C}_{33}^4[\mbox{\W$_{\{r+1,r+1\}}$}]\right)^2$ for this algebra,
even we do not
know at the moment anything about its origin.
We use the parameterization
\be
c = c_2^{r+1,r+1}(n) = \frac{(n-1) \, (r+1) \, (n -r - n r)}{n-r}
\label{11}
\en
to compute the coupling constants
of the algebras \W$_{\{r+1, r+1\}}$ along the lines of
eq.~(\ref{8}).
%and~(\ref{10}).
This leads us to the
coupling constant for $r>1$~\footnote{Eq.~(\ref{11}) is not an
allowed parameterization for $r=1$ since $c_2(2,2) = -2$ is independent of $n$.
For \W$_{\{2,2\}}$ we have instead \W$_{\{2,2\}} =$ \W$(2,3) = $ \WA$_2$; the
coupling constants for \WA$_{-2}$ are the
same as for \W$_{\infty}$.}
\be
\left({\cal C}_{33}^4[\mbox{\W$_{\{r+1,r+1\}}$}]\right)^2
\, = \,
 \frac{ \,64 \,(r+3)\,(c+2)\,(c\,(3-r) + 2 \,
(4r-3)\,(r+1))}{(r+2)\,(5c+22)\,(c\,(2-r) + (3r-2)\,(r+1))}
\label{13}
\en
and we find the surprising result that this coupling constant corresponds to
the coupling constants of the algebra \WA$_{-r-1}$ (compare
eq.~(\ref{4})).
We conclude that for $n>2$
the algebra \WA$_{-n}$ is equivalent
to the unifying algebra \W$_{\{n,n\}}$, and therefore is an algebra of
type
\W$(2,3,\ldots,n^2-1)$
Indeed, this fits nicely into
the truncations of \W$_{1+\infty}$ at negative values of the central
charge~\cite{AFMO94}.
In~\cite{FKRW94} it has been shown that \W$_{1+\infty}$
truncates to {\sf WG}$_n \cong \mbox{\sf U}(1) \oplus
\mbox{\WA}_{n-1}$ (where the Virasoro-operator is shifted such that the
{\sf U}$(1)$ current becomes a
primary spin-1 field and hence
inducing a shift by $1$ in the central charge) at a positive integer central
charge $c=n$. Extending in the above manner {\sf WG}$_n$ to negative integers
$n$ yields
algebras of type \W$(1,2,3,\ldots,(|n|+1)^2-1)$, hence in full agreement
with~\cite{AFMO94}.

This gives us a hint that
there is a strange connection between the unifying algebras
\W$_{\{r',s_o\}}$
and \W$_{\{r+s_o+1,r+2\}}$, hence between (the cosets of) the algebras
\W$^{\mbox{\sf \small sl}(r'-1)}_{\mbox{\sf \small sl}(r'-s_o)}$ and
\W$^{\mbox{\sf \small sl}(r+s_o)}_{\mbox{\sf \small sl}(s_o-1)}$ for any
fixed~$s_o$.
To be precise, we
want to put these findings in two - equivalent - statements:
\begin{itemize}
\item The coupling constants of the algebras \W$_{\{r+s_o+1, r+2\}}$
are identical
to
those of \W$_{\{r',s_o\}}$ by replacing $r' \rightarrow -r$;
\item The algebras \W$_{\{r,s_o\}}$ can be ``extended'' to negative $r$ by
defining
\W$_{\{-r,s_o\}} := \, $\W$_{\{r+s_o+1,r+2\}}$.
\end{itemize}
The proof of these statements is rather simple: Since the coupling
constants of the unifying algebras \W$_{\{r,s\}}$ depend only on the
parameterizations $c_1^{r,s}$, $c_2^{r,s}$ and $c_3^{r,s}$ (apart, of course,
{}from their form in the \WA$_{n-1}$-algebra), it is sufficient
to show that the parameterizations of W$_{\{-r,s\}}$ and \W$_{\{r+s+1,r+2\}}$
are
equivalent. So we have
\be
c_1^{-r,s} \,=\, c(n-1-r, n-1+s) \, = \,
c(n+1-(r+2), n+(r+s+1)-(r+2))\, =\, c_3^{r+s+1,r+2}
\label{14}
\en
and similarly for $c_2^{-r,s}$ and $c_3^{-r,s}$.

If we now want to know the {\sf CP}$(-k)$-model, we find the {\sf
U}$(1)$-coset of
the generalized P-B-algebra:
\be
\mbox{{\sf CP}}(-k) \,=\,
\mbox{\W}_{\{-k+2,-k+1\}} \,=\, \mbox{\W}_{\{-k+1,-k+2\}}
\,=\, \mbox{\W}_{\{(k-1)+(-k+2)+1, (k-1)+2\}} \,=\,
\mbox{\W}_{\{2,k+1\}} \, = \, \mbox{\W}_{\{k+1,2\}}\ .
\label{15}
\en
The {\sf CP}$(-2)$-model reproduces {\sf CP}$(1)$,
and {\sf CP}$(-3)$ is in this
way equivalent to the {\sf U}$(1)$-coset of the ordinary
P-B-algebra \W$(1,3/2,3/2,2)$. Therefore we
can consider the cosets of the generalized
P-B-algebras as continuations of the {\sf CP}$(k)$-models to negative
$k$.
\sect{Coset construction for \WA$_{-n-1}$}
As we mentioned, the definition~(\ref{UDS}) cannot serve for a construction for
the new algebras \WA$_{-n-1}$,
\W$_{\{n+1,n+1\}}$, and it is the only one of the unifying algebras (of \WA)
that does not fit into the general scheme
developed in ref.~\cite{BEHal94a}.
By definition,
it is an algebra with the coupling constants of \WA$_{n-1}$ where $n$
is replaced by $-n$. To find a coset construction for the algebra \WA$_{-n-1}$
we follow the arguments of ref.~\cite{BEHal94a} for the case of
\WD$_{-n}$ that we want to repeat here.
We
have to go back to the definition of classical groups of negative dimensions
of ref.~\cite{CK82}.
There it was
shown that for any scalars constructed from tensor representations the
``analytic continuation''
\be
\mbox{{\sf SO}}(-2n) \, \cong \, \overline{\mbox{{\sf Sp}}(2n)} \ \ ,
\hspace{1.5cm}
\mbox{{\sf SU}}(-n) \, \cong \, \overline{\mbox{{\sf SU}}(n)}
\label{16}
\en
holds, where the overbar means the interchange of symmetric and
antisymmetric representations. We want to give a similar relation for Kac-Moody
algebras,
where the r\^ole of the overbar takes now a transformation of the level of
the Kac-Moody:
\be
\widehat{\mbox{{\sf so}}(-2n)}_{\kappa} \,
\cong \, \widehat{\mbox{{\sf sp}}(2n)}_{\bar{\kappa}}\ \ , \hspace{1.5cm}
\widehat{\mbox{{\sf sl}}(-n)}_{\kappa} \, \cong \, \widehat{\mbox{{\sf
sl}}(n)}_{\bar{\kappa}} \  \ .
\label{17}
\en
By demanding that the Sugawara central charge
\be
c[\widehat{\cal G}_{\kappa}] \,=\, \frac{\kappa\,(\dim {\cal G})}{\kappa +
h^{\vee}}
\label{ccG}
\en
($h^{\vee}$ is the dual Coxter number)
formally agrees between the rhs
and the lhs, i.e.
\bea
c[\widehat{\mbox{\sf so}(-2n)}_{\kappa}] &=& \frac{\kappa \, (-n)\,(-2n-1)}
{\kappa + 2(-n-1)} \, = \, \frac{\bar{\kappa} \, n\,
(2n+1)}{\bar{\kappa}+(n+1)} \, = \,
c[\widehat{\mbox{\sf sp}(2n)}_{\bar{\kappa}}] \ \ ,
\label{cdn} \\[2mm]
c[\widehat{\mbox{\sf sl}(-n)}_\kappa] &=& \frac{\kappa \, (-n-1)\, (-n+1)}
{\kappa + (-n)} \, = \, \frac{\bar{\kappa} \, (n-1)(n+1)}{\bar{\kappa} + n}
\, = \,
c[\widehat{\mbox{\sf sl}(n)}_{\bar{\kappa}}]      \ \ ,
\label{can}
\ena
one can
establish the relation $\bar{\kappa} = -\kappa/2$ in the case of
$\widehat{\mbox{{\sf so}}(-2n)}$ and $\bar{\kappa} = -\kappa$ in the case of
$\widehat{\mbox{{\sf sl}}(-n)}$, respectively.

This leads us to the ansatz
of the
\W-algebras \WD$_{-n}$ and \WA$_{-n-1}$
\bea
\mbox{{\WD}}_{n} \, = \, \frac{\widehat{\mbox{{\sf so}}(2n)}_{1} \oplus
\widehat{\mbox{{\sf so}}(2n)}_{\mu}}{\widehat{\mbox{{\sf so}}(2n)}_{1+\mu}}
& \Rightarrow &
\mbox{{\WD}}_{-n} \, = \, \frac{\widehat{\mbox{{\sf sp}}(2n)}_{-1/2} \oplus
\widehat{\mbox{{\sf sp}}(2n)}_{\bar{\mu}}}
{\widehat{\mbox{{\sf sp}}(2n)}_{-1/2+\bar{\mu}}} \ \ ,\label{18}\\[2mm]
\mbox{{\WA}}_{n-1} \, = \, \frac{\widehat{\mbox{{\sf sl}}(n)}_{1} \oplus
\widehat{\mbox{{\sf sl}}(n)}_{\mu}}{\widehat{\mbox{{\sf sl}}(n)}_{1+\mu}}
& \Rightarrow &
\mbox{{\WA}}_{-n-1} \, = \, \frac{\widehat{\mbox{{\sf sl}}(n)}_{-1}
\oplus
\widehat{\mbox{{\sf sl}}(n)}_{\bar{\mu}}}
{\widehat{\mbox{{\sf sl}}(n)}_{-1+\bar{\mu}}} \ \ .
\label{19}
\ena
Eq.~(\ref{18}) is exactly the coset-construction given in~\cite{BEHal94a} for
the algebra \WD$_{-n}$. The coset~(\ref{19}) is well-defined and serves as an
definition for our new algebra \WA$_{-n-1}$.

At this point a remark of caution has to be added. One
should not take the relations~(\ref{17}) too serious: For example,
applying them to the {\sf CP}$(k)$-model one would find
\be
\mbox{{\sf CP}}(-k) = \frac{\widehat{\mbox{{\sf sl}}(-k+1)}_{\kappa}}
{\widehat{\mbox{{\sf sl}}(-k)}_{\kappa}\oplus \wh{U(1)}} \cong
\frac{\widehat{\mbox{{\sf sl}}(k-1)}_{\bar{\kappa}}}
{\widehat{\mbox{{\sf sl}}(k)}_{\bar{\kappa}}\oplus \wh{U(1)}} \ \ ,
\label{20}
\en
an expression that still doesn't make sense (for {\sf CP}$(-k)$ see
eq.~(\ref{15})).

Do we have a chance that the coset~(\ref{19}) gives indeed
an algebra of type \W$(2,3,\ldots,(n+1)^2-1)$ as demanded? It
fails already at
$n=2$: We should expect an algebra of type
\W$(2,3,\ldots,8)$; according to table I, however, there is
no primary spin-3 field
and
no truncations should take place at $\kappa = -1$
and we get an algebra of type \W$(2,4,6^2,8^2,9,10^2,12)$.
The only chance we have is that in this special case
the coset does not give rise to the \WA$_{-3}$ algebra directly but to its
orbifold. Indeed,
$\left({\cal C}_{44}^4
\right)^2$ of eq.~(\ref{scsl2cos})
with $\kappa=-1$
%\be
%\left({\cal C}_{44}^4\right)^2 \, = \,
%{{27\,{{\left( 130 + 133\,c + 5\,{c^2} \right) }^2}}\over
%   {20\,\left( 2 + c \right) \,\left( 30 + c \right) \,
%     \left( 22 + 5\,c \right) }}
%\label{21}
%\en
is identical to eq.~(\ref{4a}) with $n=-2$.
Therefore we are
confident that the coset~(\ref{19})
for $n=2$ yields the orbifold of \WA$_{-3}$ = \W$_{\{3,3\}}$.

On the other hand, for the general cosets~(\ref{cos})
with $n>2$ we have the opposite problem: these cosets have for general $\kappa$
much more fields than we need, especially for $n>3$ there are already {\em
two} primary fields of dimension~4.
To see what happens with these spin-4 primaries let us first look
at the coset $\widehat{\mbox{{\sf sl}}(n)}_\kappa/\mbox{{\sf sl}}(n)$,
following the calculations of~\cite{BBSS88}.
%\be
%W_3 \,=\, j_a q_a
%\label{23}
%\en
%with
%\be
%q_a \,= \, d_{abc}\, j_b j_c
%\label{24}
%\en
%It is easy to show that $W_3$ is primary under the stress-energy tensor
%\be
%T \, = \,\frac{1}{2\,(\kappa+n)}\, j_a j_a
%\label{25}
%\en
Apart from $W_3'$, there are four invariant fields at dimension~4,
$T''$, $TT$ and two independent fields, of which we can construct the two
primary fields.
We computed the Kac-determinant of these invariants for general
$n$ and $\kappa$ and obtain
\be
\det {\cal M}_4 \, \sim \,
\frac{\kappa^4 \, (\kappa^2-1) \, (n^2-9) \,(n^2-4) \, (n^2-1)^4 \,
    (2 \kappa + n) \,(3 \kappa + 2n)}{(\kappa + n)^2} \ \ .
\label{26}
\en
This shows the well-known fact~\cite{BBSS88} that there is a null-field for
general $n$ at $\kappa=1$, but there is also a null-field at $\kappa=-1$.
For $n=3, \kappa =1$ both primary spin-4 fields become null-fields and the
coset truncates to the algebra \WA$_{2}$, whereas
for $n=3, \kappa=-1$ one of these fields survives.

Let us now turn back to the general coset~(\ref{cos}).
We know~\cite{BBSS88a} that these cosets truncate to the \WA$_{n-1}$-algebras
if we
set
$\kappa=1$. For the
general coset we computed the coupling constant ${\cal C}_{33}^4$ (here the
spin-4 field is taken to be the one that appears in the OPE of $W_3$ with
itself) and obtain
(expressing $\mu$ as function of $n$, $\kappa$ and the central charge~$c$)
\bea
{\cal C}_{33}^4(n,\kappa,c) &=& 64 \,
\left[{c^2}\,\left( -3 + n \right) \,\left( 3 + n \right) \,
   \left( {\it \kappa} + n \right) \,{{\left( 2\,{\it \kappa} + n \right) }^2}
+  \right. \nonumber \\[1mm]
&&
  c\,{n^2}\,\left( -23\,{\it \kappa} - 73\,{{{\it \kappa}}^3} - 14\,n -
     146\,{{{\it \kappa}}^2}\,n - 65\,{\it \kappa}\,{n^2} +
     9\,{{{\it \kappa}}^3}\,{n^2} - 2\,{n^3} + 18\,{{{\it \kappa}}^2}\,{n^3} +
     8\,{\it \kappa}\,{n^4} \right)
+ \nonumber \\[1mm]
&&
\left.
  2\,{\it \kappa}\,\left( -1 + n \right) \,\left( 1 + n \right) \,
   \left( -72\,{{{\it \kappa}}^2} - 144\,{\it \kappa}\,n - 65\,{n^2} +
     7\,{{{\it \kappa}}^2}\,{n^2} + 14\,{\it \kappa}\,{n^3} + 8\,{n^4} \right)
\vs{2}
\right]\,\times  \nonumber \\[1mm]
&&
\left[
\left( 22 + 5\,c \right) \,\left( -2 + n \right) \,\left( 2 + n \right) \,
  \left( 2\,{\it \kappa} + n \right)\vs{2}
 \, \right. \nonumber \\[1mm]
&&
\left.
\vs{2}  \left( c\,\left( {\it \kappa} + n \right) \,\left( 2\,{\it \kappa} + n
\right)  +
    {\it \kappa}\,\left( -1 + n \right) \,\left( 1 + n \right) \,
     \left( 2\,{\it \kappa} + 3\,n \right)  \right) \right]^{-1}
\label{27}
\ena
For $\kappa=1$ the coupling constant~(\ref{27}) is the same as
eq.~(\ref{4}) of \WA$_{n-1}$. For us it is important to realize
that it is invariant
under the transformation
\be
n\rightarrow -n, \hspace{1.5cm} \kappa \rightarrow -\kappa \ \ .
\label{28}
\en
This
leads to the symmetry of the coupling constant ${\cal C}_{33}^4(-n,\kappa,c)
= {\cal C}_{33}^4(n,-\kappa,c)$, i.e.\ it supports the conjecture~(\ref{19})
for the algebra \WA$_{-n-1}$.
When our ansatz
for a coset-construction for \WA$_{-n-1}$
is true, then all coupling constants ${\cal C}(n,\kappa,c)$ of the
coset~(\ref{cos}) should have this symmetry.
One has also to show that this coset with $\kappa=-1$ is indeed an algebra of
the indicated type. This would conclude the proof of the existence and the
exact definition of
\WA$_{-n-1}$. Still missing, however, is a coset construction for \WA$_{-3}= $
\W$(2,3,\ldots,8)$.

A different starting point, that of pseudo-differential operators, has been
chosen in refs.~\cite{KZ93,KM94} to investigate the
{\em classical} algebras \WA$_{n-1}$ for {\em complex} $n$.
We should note, however, that
contrary to the quantum algebra we
do not expect a truncation of \WA$_{n-1}$ for negative $n$ to a finitely
generated algebra in the classical case. It will be very interesting to see
these two approaches converge.\vs{10}

I would like to thank the University of Torino, Department of Physics, for kind
hospitality and A.~Honecker for
reading the manuscript. Moreover, I am grateful to L.~Feh\'er for drawing my
attention to
refs.~\cite{KZ93} and~\cite{KM94}.

\renewcommand{\baselinestretch}{0.6}

{\small
\begin{thebibliography}{9}
\bibitem{FL90} V.A.~Fateev and S.L.~Luk'yanov, {\em Additional Symmetries and
Exactly Solvable Models in Two-Dimensional Conformal Field Theory}, Sov.\ Sci.\
Rev.\ A.~Phys. {\bf 15/2} (1990)
\bibitem{BS93} P.~Bouwknegt and
K.~Schoutens, {\em \W-Symmetry in Conformal Field
Theory}, \PR{223} (1993) 183
\bibitem{Hor94} K.~Hornfeck, {\em Classification of
Coupling Constants for \W-algebras from
Highest Weights}, \NP{B411} (1994) 307
\bibitem{KW91} H.G.~Kausch and~G.M.T~Watts, {\em A Study of \W-algebras Using
Jacobi-Identities}, \NP{B354} (1991) 740
\bibitem{LF89} S.L.~Luk'yanov and V.A. Fateev, {\em Exactly soluable models of
conformal quantum field theory associated with the simple Lie algebra {\sf
D}$_n$}, Sov.~J.~Nucl.~Phys.~{\bf 49} (1989) 925
\bibitem{EHH93} W.~Eholzer, A.~Honecker and R.~H\"ubel, {\em How Complete is
the Classification of \W-algebras?}, \PL{B308} (1993) 42
\bibitem{BEHal94a}R. Blumenhagen, W. Eholzer, A. Honecker, K. Hornfeck and R.
H\"ubel, {\em Coset-Realizations of Unifying \W-algebras}, preprint
DFTT-25/94, BONN-TH-94-11, hep-th/9406203
\bibitem{BEHal94}R. Blumenhagen, W. Eholzer, A. Honecker, K. Hornfeck and R.
H\"ubel, {\em Unifying \W-algebras}, \PL{B332} (1994) 51
\bibitem{BK90} I. Bakas and E. Kiritsis, {\em Universal \W-algebras in quantum
field theory}, \IJMP{A6} (1990) 2871
\bibitem{Blu91} R. Blumenhagen, {\em \W-algebras in conformal quantum field
theory}, Diplomarbeit, BONN-IR-91-06
\bibitem{Vos92} K. de Vos, {\em Coset algebras, integrable hierarchies and
matrix
models}, Ph.D.~Thesis (1992)
\bibitem{BBSS88a} F.A.~Bais, P.~Bouwknegt, M.~Surridge and K.~Schoutens, {\em
Coset construction for extended Virasoro algebras},
\NP{B304} (1988) 371
\bibitem{Hor92} K.~Hornfeck, {\em The Minimal Supersymmetric Extension of
\WA$_{n-1}$}, \PL{B275} (1992) 355
\bibitem{BG88} P.~Bowcock, P.~Goddard, {\em Coset Constructions and Extended
Conformal Algebras}, \NP{B305} (1988) 685
\bibitem{Alt89} D.~Altschuler, {\em Quantum Equivalence of Coset Space Models},
\NP{B313} (1989) 293
\bibitem{BK90a} I. Bakas and E. Kiritsis, {\em Grassmanian Coset Model and
Unitary Representations of \W$_{\infty}$}, \MPL{A8} (1990) 2039
\bibitem{Nar91} F.J. Narganes-Quijano, {\em On the parafermionic \W$_N$
Algebra},
\IJMP{A6} (1991) 2611
\bibitem{BK92} I. Bakas and E. Kiritsis, {\em Beyond the Large $N$ Limit:
Non-Linear \W$_{\infty}$ as Symmetry of the {\sf sl}$(2)$/{\sf U}$(1)$ Coset
Model}, \IJMP{A7}, Suppl.~1A (1992) 55
\bibitem{BFH94} J.~de~Boer, L.~Feh\'er and A.~Honecker, {\em A Class of
\W-Algebras with Infinitely Generated Classical Limit}, \NP{B420} (1994) 409
\bibitem{Rag94} E.~Ragoucy, private communication
\bibitem{Hor93} K.~Hornfeck, {\em \W-algebras with set of primary fields of
dimensions~(3,4,5) and~(3,4,5,6)}, \NP{B407} (1993) 237
\bibitem{AFMO94} H.~Awata, M.~Fukuma, Y.~Matsuo, S.~Odake, {\em Character and
Determinant Formulae of Quasi-Finite Representations of \W$_{1+\infty}$
Algebra}, preprint YITP/K-1060, YITP/U-94-17, SULDP-1994-3, hep-th/9405093
\bibitem{FKRW94} E.~Frenkel, V.~Kac, A.~Radul, W.~Wang, {\em \W$_{1+\infty}$
and \W({\sf gl}$_N$) with central charge N}, preprint hep-th/9405121
\bibitem{CK82}P.~Cvitanovi\'c and A.D.~Kennedy, {\em Spinors in Negative
Dimensions}, Physica Scripta {\bf 26} (1982) 5
\bibitem{BBSS88} F.A.~Bais, P.~Bouwknegt, M.~Surridge and K.~Schoutens, {\em
Extensions of the Virasoro algebra constructed from Kac-Moody algebras using
higher order Casimir invariants}, \NP{B304} (1988) 348
\bibitem{KZ93}B.~Khesin and I.~Zakharevich, {\em Poisson-Lie Group of
Pseudodifferential Symbols}, preprint hep-th/9312088, to appear in Comm.\
Math.\ Phys.
\bibitem{KM94}B.~Khesin and F.~Malikov, {\em Universal Drinfeld-Sokolov
Reduction and Matrices of Complex Size}, preprint hep-th/9405116
\end{thebibliography}
}
\end{document}